\documentstyle[11pt,psfig]{article}
\begin{document}
\centerline{  \LARGE \bf Chiral symmetry restoration in excited hadrons.}
\bigskip
\medskip
\centerline{ \Large L. Ya. Glozman}
\medskip
\centerline{\it  Institute for Theoretical
Physics, University of Graz, Universit\"atsplatz 5, A-8010 Graz,}
\centerline{\it Austria}

\vspace{0.7cm}
\noindent
{\bf Abstract.} 
The evidence, theoretical justification and implications
of chiral symmetry restoration in excited hadrons are presented.
\vspace{0.7cm}

\bigskip
\bigskip
\large
Different 3Q potential constituent quark models \cite{CI,GPVW,M}
predict a lot of baryon states at the mass region 2 GeV, which
are not observed. A possible discovery of the strange pentaquark
suggests that there should be in addition high-lying 5Q (and with
higher amount of quarks) states that belong to different multiplets.
Both constituent quark models as well as different pentaquark
models rely crucially on spontaneous breaking of chiral symmetry
(the very notion of constituent quark as a quasiparticle is
related to chiral symmetry breaking in the QCD vacuum). A 
question then arises why all these states are not seen? Is it
related to technical limitations to see them or it perhaps
indicates that the physics of the high-lying hadrons is different
as compared to the low-lying ones? Indeed, if one looks carefully
at the nucleon excited states, see Fig. 1,
 one immediately
notices regularities for high-lying states that definitely
absent for the low-lying states. In particular, the nucleon (and delta)
high-lying spectra show obvious patterns of parity doubling:
starting from the 1.7 GeV region excited nucleons of the same spin
but opposite parity are approximately degenerate. Is it accidental?
If not, some symmetry should be behind this parity doubling.
Then we have to understand why this symmetry is definitely
absent for the low-lying states and persists only higher
in the spectrum.

\begin{figure}
\hspace*{-0.5cm}
\centerline{
\psfig{file=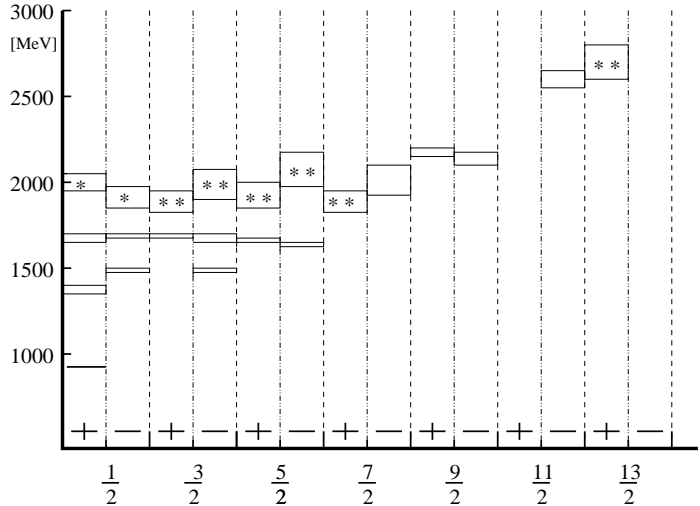,angle=-90,width=0.6\textwidth}
}
\caption{ \large Excitation spectrum of the nucleon. The real part
of the pole position is shown. Boxes represent experimental
uncertainties. Those resonances which are not yet established
are marked by two or one stars according to the PDG classification.
The one-star resonances with $J=1/2$ around 2 GeV are given
according to the recent Bonn (SAPHIR) results. }
\end{figure}

\medskip
It has been suggested some time ago that this parity doubling
reflects effective chiral symmetry restoration \cite{G1}. We
know that the QCD Lagrangian posseses almost perfect
$SU(2)_L \times SU(2)_R$ chiral symmetry, which is a symmetry
of two independent (left and right) rotations of $u$ and $d$
quarks in the isospin space. If this symmetry of the QCD
Lagrangian was intact in the QCD vacuum, then all hadrons
would fall into representations of the parity-chiral group
$SU(2)_L \times SU(2)_R \times C_i$, where the group $C_i$
consists of elements identity and space inversion \cite{CG2}.
For baryons these multiplets are either parity doublets in
$N$ and $\Delta$ spectra, not related to each other, or quartets
that contain degenerate nucleon and delta doublets. Either
of these degeneracies should be fulfilled for any spin of baryons,
because chiral symmetry of QCD does not constrain spin. From
the low-lying nucleon and delta spectra we can conclude that
there are no degeneracies of states of the same spin but
opposite parity (even more, there is  no one-to-one mapping
of states of opposite parity with the same spin). It is this
fact which was historically one of the most important arguments
to conclude that chiral symmetry must be spontaneously broken.
This spontaneous breaking leads to the appearance of quasiparticles
(constituent quarks) and gives a basis for a constituent quark
model. The latter one, supplemented by the effective interactions
of constituent quarks which are also related to chiral symmetry
spontaneous breaking \cite{GR,GPVW}, is known to explain
regularities of the low-lying baryons. However, higher in the
spectrum the typical momenta of constituent quarks should increase,
consequently the quasiparticle (constituent) mass of quarks
should drop off and chiral symmetry will be effectively restored
\cite{G1}. This is a plausible microscopical picture of symmetry
restoration.

\medskip
The systematic approach to the symmetry restoration based on
QCD has been formulated in ref. \cite{CG1,CG2}. By definition
an effective symmetry restoration means the following. In QCD
the hadrons with the quantum numbers $\alpha$ are creared when
one applies the local interpolating field (current) $J_\alpha$
with such quantum numbers on the vacuum $|0\rangle$. Then all
the hadrons that are created by the given interpolator appear
as intermediate states in the two-point correlator

\begin{equation}
\Pi =\imath \int d^4x ~e^{\imath q x}
\langle 0 | T \{ J_\alpha (x) J_\alpha (0) \} |0\rangle,
\label{corr}
\end{equation}

\noindent
where all possible Lorentz and Dirac indices (specific for
a given interpolating field) have been omitted. Consider
two local interpolating fields $J_1(x)$ and $J_2(x)$ which
are connected by chiral transformation, $J_1(x) = UJ_2(x)U^\dagger$,
where $U$ is an element of the chiral group. Then, if the vacuum
was invariant under chiral group, $U|0\rangle = |0\rangle$,
it follows from (\ref{corr}) that the spectra created by
the operators $J_1(x)$ and $J_2(x)$ would be identical. We know
that in QCD one finds $U|0\rangle \neq |0\rangle$. As a consequence
the spectra of two operators must be different. However, it may
happen that the noninvariance of the vacuum becomes unimportant
(irrelevant) high in the spectrum. Then the spectra of both
operators become close al large masses (and asymptotically
identical). This would mean that chiral symmetry is effectively
restored. We stress that this effective chiral symmetry
restoration does not mean that chiral symmetry breaking in
the vacuum disappears, but only that the role of the quark
condensates that break chiral symmetry in the vacuum becomes
progressively less important high in the spectrum \cite{CG1,CG2}.
One could say, that the valence quarks in high-lying
hadrons decouple from the QCD vacuum.

\medskip
Actually it is easy to prove that it must happen in QCD. At
large space-like momenta $Q^2 = -q^2 $ the correlator can be 
adequately represented by the operator product expansion,
where all nonperturbative effects reside in different
condensates \cite{SVZ}. The only effect that spontaneous
breaking of chiral symmetry can have on the correlator is
via the quark condensates of the vacuum. However, the
contributions of all these condensates are suppressed
by inverse powers of $Q^2$. This shows that even if the
chiral symmetry is broken in the vacuum, at large space-like
momenta the correlation function becomes chirally symmetric.
In other words $\Pi_{J_1}(Q) \rightarrow  \Pi_{J_2}(Q)$ at
$Q^2 \rightarrow \infty$. The dispersion relation provides
a connection between the space-like and time-like domains
of the correlator. In particular, the large $Q^2$ correlator
is completely dominated by the large $s$ spectral density.
Hence the spectral density at large $s$ should be insensitive
to the chiral symmetry breaking in the vacuum and must satisfy
$\rho_{1}(s) \rightarrow  \rho_{2}(s)$ at
$s \rightarrow \infty$. If this chiral symmetry restoration
happens in the regime where the spectrum is still
quasidiscrete (i.e. it is dominated by resonances and the
successive resonances with the given spin are separated),
then these resonances must fill in representations of the
parity-chiral group.

\medskip
Clearly it is a matter of experiment to answer a question
at which mass scale it happens. For example, the difference
between the vector and axial-vector spectral densities,
extracted from the weak decays of tau-lepton by ALEPH \cite{ALEPH} and
OPAL \cite{OPAL} collaborations is compatible with zero at
masses of 1.7 GeV, though uncertainties are rather large. This
difference is entirely from the spontaneous breaking of chiral
symmetry and while it is large at $\rho$ and $a_1$ masses, it
gets strongly suppressed at the same mass scale where we see
parity doublets in Fig. 1

\medskip
A direct evidence \cite{G2} for chiral symmetry restoration
can be infered from the recent results of partial wave
analysis of proton-antiproton annihilation at LEAR \cite{BUGG}.
In particular, the pions, $I,J^P =1,0^-$, and $\frac{1}{\sqrt {2}}(u\bar u + d\bar d)$ $f_0$ mesons, $I,J^P =0,0^+$ would form  $(1/2,1/2)$
representations of the chiral group and would be degenerate level
by level if the vacuum was chirally symmetric. Clearly it is
not the case for the low-lying mesons, where effects of chiral
symmetry breaking in the vacuum are strong. However, starting
from 1.5 GeV mass we observe a pattern of chiral
symmetry restoration, see Fig. 2.

\medskip
The phenomenon of chiral symmetry restoration in hadron spectra
rules out the nonrelativistic potential description of
high-lying hadrons in the spirit of the constituent quark
model \cite{G2,G3,G4}. Consider, for instance, mesons. Within
the nonrelativistic potential description of mesons the parity
of the state is unambiguously prescribed by the relative orbital 
momentum of quarks. The states of opposite parity require
different orbital angular momenta and hence different centrifugal
repulsion. Hence they cannot be systematically
degenerate within such a picture.
Similar conclusions can be obtained for baryons \cite{G1}. Clearly
the chiral symmetry restoration in high-lying hadrons is
against relativistic scalar confinement potential description either, 
because scalar confinement manifestly breaks chiral symmetry. However,
it does not contradict Lorentz-vector confining potential and
this type of potential can be reconciled with parity doubling \cite{L}.

\begin{figure}
\hspace*{-0.5cm}
\centerline{
\psfig{file=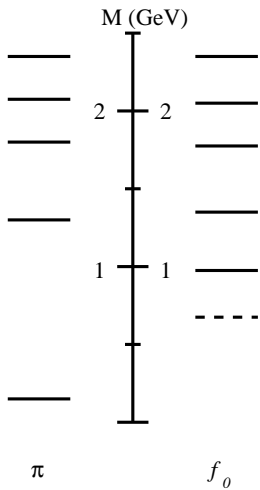,angle=-90,width=0.6\textwidth}
}
\caption{\large Pion and $n \bar n$ $f_0$ spectra. }
\end{figure}

\medskip
The most simple model of highly-excited hadrons compatible with the
chiral symmetry restoration is a string (with the color-electric flux-tube
in the string) and with bare quarks of definite chirality at the
ends of the string \cite{G3}. Once the quarks posess a definite
chirality, then all hadrons will form degenerate multiplets
of opposite parity. For example, $n\bar n$ $f_0$ mesons and pions 
 represent the following combinations of the right and left quarks

\begin{equation}
f_0: \frac{1}{\sqrt 2} (\bar R  L +  \bar L  R),
\end{equation}

\noindent

\begin{equation}
\pi : \frac{1}{\sqrt 2} (\bar R  \tau L - 
\bar L  \tau R).
\end{equation}

Similar valence quark content relations can be written
for other hadrons. Actually, parity partners represent
different parity states of the same "basic" particle, energy
of which is determined by the energy of the string.
As a byproduct, this type of model automatically solves a famous
spin-orbit problem of the constituent quark model: Since the
chirality operator does not commute with the spin-orbit
operator, there is no spin-orbit force once the chiral symmetry
is restored.

\medskip
Similar arguments about chiral symmetry restoration in meson
spectra have been suggested in ref. \cite{B,S}. It has also been
shown in ref. \cite{O} that the string spectrum can be reproduced
via the Salpeter-type or Dirac equations with vector confinement.
Very interesting evidence for the chiral symmetry restoration 
has been obtained
in lattice calculations \cite{D}, where it has been shown that
the high-lying mesons decouple from the low-lying eigenmodes
of the Dirac operator (which determine the quark condensate).

\medskip
Clearly, the systematic experimental study of high-lying hadrons will be
an interesting enterprise and will allow us to understand a lot
about QCD in the confining regime. This program is just on the way
at ELSA, JLAB and hopefully will be  important for the future
 antiproton ring at GSI and for a new Japanese hadron
facility.

\end{document}